\input{psfig.sty}
\documentstyle[12pt]{article}
\newfont{\feff}{cmti10}
\begin{document}

\title{On a puzzle  posed by 
 the Kurien-Sreenivasan  boundary layer experiment. }

\author{ Victor Yakhot\\
Department of Aerospace and Mechanical Engineering\\
Boston University, 
Boston, MA 02215 }

\maketitle

${\bf Abstract}.$
\noindent
The Kolmogorov  inertial range 
ratio of the mixed- to- longitudinal third-order structure functions 
 of isotropic and homogeneous turbulence  
is $\frac{S_{1,2}}{S_{3,0}}=1/3$. Recent measurements
by Kurien and Sreenivasan ( Phys. Rev. E {\bf 64}, 056302 (2001)), showed that,  while the 
longitudinal structure function was extremely close to Kolmogorov's, 
the measured  ratio  was
very far:
$\frac{S_{1,2}}{S_{3,0}}\approx 0.43$. 
Explanation of  this puzzle is presented in this letter. 

\newpage

\noindent Kolmogorov's  relations  for the third-order structure functions 
in isotropic
and homogeneous turbulence, derived in 1941 for a three-dimensional flow, are:

\begin{equation}
S_{3,0}\equiv \overline{(\Delta u)^{3}}=\overline{(u(x+r)-u(x))^{3}}
\approx (-1)^{d}\frac{12}{d(d+2)}{\cal E}r
\end{equation}

\noindent and 

\begin{equation}
S_{1,2}\equiv \overline{\Delta u (\Delta v)^{2}}=\frac{S_{3,0}}{3}
\end{equation}

\noindent where $u$ and  $v$ are  components of the velocity field 
 in the $x$ and $y$-directions, respectively and $r$ is a displacement ,
chosen parallel to the $x$-axis.
The mean dissipation rate ${\cal
E}=\nu\overline{(\partial_{i}v_{j})^{2}}$ and $d=2;~3;$ is  space
dimensionality, This relation is important since, being  exact in the
inertial range (large Reynolds number), it enables one to test experimental
conditions and quality of the flow, 
define inertial ranges and even obtain exponents of the high order
structure functions. 

\noindent Recent high Reynolds number measurements,  performed in the boundary layer
[2], 
showed that,  while the data on $S_{3,0}$ were 
 very close to the relation (1), the measured mixed
correlation function $S_{1,2}\approx 0.43 S_{3,0}$ strongly contradicted the
exact relation (2). At the same time , the measurements of Ref.  [2] gave 
$S_{0.2}/S_{2,0}\approx 4/3$, very close to the exact relation of the theory
of isotropic and homogeneous turbulence.  This situation is  puzzling,  
since it indicates that the
flow can simulataniously 
be both isotropic (relation (1) for $S_{3,0}$) and inisotropic (relation (2)
for $S_{1,2}$).  

\noindent To address this problem, we use the 
 set of exact relations  for the inertial range of homogeneous
turbulence,  derived in Ref. [3]. Introduce a generating function 

$$Z(\eta_{2},\eta_{3})=<exp({\eta_{2}\Delta u +\eta_{3}\Delta v})>$$

\noindent so that $S_{n,m}=\overline{(\Delta u)^{n}(\Delta
v)^{m}}=\partial_{2}^{n}\partial_{3}^{m} Z(0,0)$. As we see, given the equation
for the generating function, derived in Ref. [3], the relations for 
all structure
functions can be obtained in this simple manner. The exat equations  for
$S_{n,m}$ were 
applied to explanation of the so called extended self-
similarity in Ref.[ 4].

\noindent The relations for the third-order moments [3] are:

\begin{equation}\frac{\partial S_{1,2}}{\partial r}+\frac{d+1}{r}S_{1,2}=
-2<(\Delta
\frac{\partial p}{\partial y})\Delta v>+(-1)^{d}\frac{4}{d}{\cal E} 
\end{equation}

\noindent and

\begin{equation}\frac{\partial S_{3,0}}{\partial
r}+\frac{d-1}{r}S_{3,0}-2\frac{d-1}{r}S_{1,2}
=-2<(\Delta
\frac{\partial p}{\partial x})\Delta u>+(-1)^{d}\frac{4}{d}{\cal E} 
\end{equation}

\begin{equation}
\frac{\partial S_{2,0}}{\partial
r}+\frac{d-1}{r}S_{2,0}-\frac{d-1}{r}S_{0.2}=0
\end{equation}

\noindent where $\Delta \partial_{i}p=\partial_{i}p(x+r)-\partial_{i}p(x)$. 
The relation (5),  valid for any divergence-free statistically isotropic
field, is kinematic  and,  as such, it 
 does not include pressure terms, which are a feature of hydrodynamic
equations. This relation 
  gives the 
well-known relation $S_{2,0}/S_{0,2}\approx 4/3$, provided $S_{2,0}\propto
S_{0,2}\propto r^{\xi_{2,0}}$ with $\xi_{2,0}\approx 2/3$. 

The dynamics of  the higher-order velocity correlation functions is strongly
influenced by the
pressure terms and thus, we will try to understand   the deviations 
from the
Kolmogorov law (1),(2) observed in Ref.[2] 
as originating from  the boundary layer flow 
pecularities 
of  the pressure contributions to 
(3), (4). 

In the isotropic and homogeneous flow,  the pressure contributions 
are equal to
zero since $\overline{v_{i}(x)\partial_{i}p(x+r)}=\overline{
\partial_{i}v_{i}(x)p(x+r)}=0$ for any value
of displacement $r$ , even $r=0$. Since the
flow is isotropic, both $\overline{u(x)\partial_{x}
p(x+r)}=\overline{v(x)\partial_{y}p(x+r)}=0$. 
Solving equations (3) and (4) gives the
Kolmogorov relations (1), (2),

 The velocity field in a boundary layer is $(U+u){\bf i}+(V+v){\bf
j}+(W+w){\bf k}$, where  the units vectors ${\bf i}$, ${\bf j}$ and ${\bf k}$
are  parallel to 
the stream-wise  ( along $x$-axis),
transverse (perpendicular to the wall; y-axis) and spane-wise 
(z-axis) directions, respectively. The  
capital and small letters denote the  mean and fluctuating (turbulent)
contrubutions to the velocity field, respectively.
 Since $W\approx 0$,   
the mean velocity distribution 
$\overline{\bf v}=U(y){\bf i}+V{\bf j}$ , and  $\overline{uv}\propto
\partial_{y}U(y)$ where $y$ is the distance to the wall. The   the turbulent 
kinetic energy is mainly pumped into
the $u$-component of velocity field  and  
the role of the pressure fluctuations
is to redistribute this energy between velocity components. Thus,
the pressure terms serve as an effective energy  source in the equation for
 $\overline{v^{2}}$ and as a sink in the equation for $\overline{u^{2}}$,
leading to the ``return -to-isotropy'' process.
This means that if we introduce a measure of anisotropy: 
$\overline{v(x)\partial_{y}p(x+r)}=
(-1)^{d+1}\kappa/2;~\overline{u(x)\partial_{x}p(x+r)}=(-1)^{d+1}\xi/2$
with both $\kappa$ and $\xi=O(1)$, then $\xi<0$ , while $\kappa>0$. 
This conclusion is quite plausable: the pressure terms generate both
$v$-component of velocity field and all its correlations furnctions.

\noindent In the boundary layer $\partial_{y}v>>\partial_{z}w$
and,  since $p(x)(\partial_{y}v+\partial_{x}u)\approx 0$, 
 we assume  the deviations from the
isotropic pressure contributions (equal to zero) 
to be two -dimensional, generating only two large 
pressure-velocity correlations 
$\kappa$ and $\xi$. If this is so, then $\kappa\approx -\xi$.

\noindent From the equations (3) and (4) we have, 

\begin{equation}
S_{1,2}=(-1)^{d}\frac{4}{d(d+2)}({\cal E}+\frac{4}{d}\kappa)r
\end{equation}

\noindent and 

\begin{equation}
S_{3,0}=(-1)^{d}[\frac{12}{d(d+2)}{\cal E}+
\frac{1}{d}(\xi+\frac{2(d-1)}{d+2}\kappa)]r
\end{equation}

\noindent  It was proposed  in [3] that one of the measures of the flow 
anisotropy is the
third-order structure function of  transverse components of velocity field
$S_{0,3}=\overline{(\Delta v)^{3}}>$, which is equal to zero
 in  isotropic turbulence. The
measured value [2], however, was $S_{0,3}\approx 0.1 S_{3,0}$. Using this as an
empirical estimate, chracterizing the anisotropic effects, we see that if 
 $\kappa\approx 0.2{\cal E}$ and , as was argued above,
$\kappa\approx -\xi$, we obtain $S_{3,0}=-0.79{\cal E}r$ and and
$S_{1,2}/S_{3,0}\approx 0.43$, very close to the experimental data [2]. 

The  above estimates   justify  using the isotropic relations (3) and (4) to
analyze this weakly anisotropic flow:  the magnitudes of the 
pressure -generated 
``sources'' and ``sinks'' $\kappa$ and $\xi$ are only a relatively 
small fraction of the 
corresponding dissipation rate ${\cal E}$ and thus, the pressure contributions
can be treated as small perturbations to the isotropic relations of Ref.[3]. 
The ``large'' observed deviation from the 
isotropic ratio (2) 
is partially  due to a large 
 numerical factor  in the relation (6).   The most
interesting outcome of the present work is that the  cancellation in (7),
makes the deviation form K41 expression for 
$S_{3,0}$ practically  non-observable.  

\noindent It has been shown in  Ref. [3] (see also [2]) , 
that the the equations for  the
mixed odd-order structure functions $S_{1,2n}$ are ``simple'': in addition to
the pressure and dissipation terms, they involve the $S_{1,2n}$ functions
only. At the same time, the equations for the longitudinal structure functions 
$S_{2n+1,0}$ are ``mixed'' with explicit  contributions from $S_{1,2n}$ , thus
enabling the cancellations, leading to close- to- isotropic results which can be
tested against numerical simulations.   The conclusions   of this paper rely on
some features of the boundary layer flows. It is hard to say how
universal the mechanism is. 

These results can be interpreted  in a following way. In the limit
$\nu_{0}\rightarrow 0$ , the Fourier -transforms of Kolmogorov relations,
involving analytic functions only,  are 
equal to zero for $k\neq 0$, reflecting the constancy of the energy flux,
originating at $k=0$. The linear in $r$ pressure-velocity correlations,
originating at the anisotropic large scales, equal to zero in the $k\neq 0$
 interval, 
cannot and must not disappear in the inertial range. That is why in these
flows one cannot
observe  the expected ``return-to-isotropy''  in the
Kolmogorov relations. This fact was amply demonstrated by the
Kurien-Sreenivasan experiment [2].
The non-analytic even-order structure functions, having
non-zero Fourier components in the inertial range,  are another matter: the
large-scale (enery range) contributions can be readily ``forgotten''  in the
inertial range. This may explain why it is so hard to experimentally 
observe the predicted 
relations (1),(2) for the third order structure functions even in the very high
Reynolds number flows.

\noindent {\bf references}\\
1.~A. N. Kolmogorov, Dokl. Akad. Nauk. SSSR, {\bf 32}, 19 (1941).\\
2.~ S. Kurien and K. R. Sreenivasan, Phys. Rev.E., {\bf 64}, 056302 (2001)\\
3.~ V. Yakhot, Phys. Rev. E., {\bf 63}, 026307 (2001).\\
4. V. Yakhot, Phys. Rev. Lett., {\bf 83}, 234501 (2001)\\ 
\\

\end{document}